\documentclass[numbers,preprint, 12pt]{elsarticle}

\usepackage{epsfig}

\usepackage{amssymb}

\journal{NIMBioS}

\begin{document}

\begin{frontmatter}



\title{Modeling the effects of cymene on the distribution of germination and growth of \emph{Beauveria bassiana}}


\author[mhc]{Luong Nguyen} 
\address[mhc]{Department of Mathematics, Mount Holyoke College, 50 College Street, South Hadley, MA 01075} 

\author[hood]{Dubravka Bodiroga}
\address[hood]{Department of Mathematics and Computer Science, Hood College, Frederick, MD 21701}

\author[iastate]{Reka Kelemen}
\address[iastate]{ Department of Genetics, Development and Cell Biology, Iowa State University,, IA 50011-3223}

\author[utk2]{Jaewook Joo}
\ead{jjoo1@utk.edu}
\address[utk2]{Department of Physics and Astronomy,The University of Tennessee, Knoxville, TN 37996}

\author[utk1]{Kimberly D. Gwinn}
\ead{kgwinn@ut.edu}
\address[utk1]{Department of Entomology and Plant Pathology, The University of Tennessee, Knoxville, TN 37996}

\begin{abstract}
Essential oils have antifungal and antipathogenic effects and therefore are targets in plant pathology research for their potential uses as natural substitutes for inorganic plant pesticides. \emph{Beauveria bassiana}, an entomopathogenic fungus, can endophytically colonize a vast number of plant species and trigger induced systemic resistance against plant pathogens. Spore germination is the most vulnerable in the fungal life cycle and is therefore a good candidate for monitoring the effect of essential oils on the growth of \emph{B. bassiana}. Percentage germination of fungal spores and length of germination tubes were recorded from experiments. A mathematical model that was able to
capture the effects of cymene, an essential oil produced by Monarda, on the germination and growth was developed. This is the first report of a model for the impact of essential oils on \emph{B. bassiana} spore germination.
\end{abstract}


\end{frontmatter}

\section{Introduction}

\emph{Monarda}, a member of the family \emph{Lamiaceae} native to the Great Smoky Mountains National Park, synthesizes highly volatile substances, called essential oils. Essential oils have an antifungal and antipathogenic effect and therefore are targets in plant pathology research because of their potential as natural substitutes for inorganic plant pesticides. 

The role of essential oils in control of \emph{Rhizoctonia} damping-off in tomato plant with bioactive \emph{Monarda} herbage was the focus of the study conducted at the University of Tennessee in Knoxville. Tomato seeds grown in a medium containing essential oils extracted from \emph{Monarda} were protected against damping-off disease caused by fungus \emph{Rhizoctonia} solani [1]. \emph{Beauveria bassiana}, a beneficial, entomopathogenic fungus, can endophytically colonize a vast number of plant species and can also trigger induced systemic resistance against plant pathogens [2].The overall goal of this project was to determine if the combination of the treatment of tomato seeds with bioactive herbage and with \emph{B. bassiana} enhances the resistance of the tomato plant against its fungal and bacterial pathogens. 

A spore germination of \emph{B. bassiana} occurs after a spore dispersed from a colony lands on a moist surface and is activated by conditions that are optimal for fungal growth, such as heat, moisture and nutrients (Fig. 1). As water molecules from the environment enter the spore, the spore starts to swell; its size increases until it reaches a critical volume when a small protrusion starts to form on one side of the spore. This protrusion will develop into a branching germ tube that will develop into the fungal hyphae. This initial stage of fungal growth is the most vulnerable in the fungal life cycle and is therefore a good candidate for monitoring the effect of essential oils on the growth of \emph{B. bassiana}. 

The overall aim of this study was to model the effect of cymene, an essential oil produced by \emph{Monarda}, on the germination and growth of the beneficial fungus, \emph{B. bassiana}. Our specific objectives were to measure the spore germination percentage and model the spore germination tube growth rate as a function of cymene concentration. 

\begin{figure}
\centering
\includegraphics[width=2in]{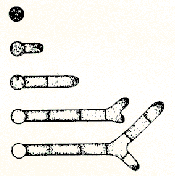}
\caption{Fungal spore germination depicting the stages of spore germination formation and branching}
\label{germStages}
\end{figure}

\section{Materials and Methods}
Combinations of cymene and ethanol were used to produce six different concentrations of cymene: 500, 50, 5, 0.5, 0.05, and 0.005 mM. Ethanol was used as the control. A 200 $\mu$l capacity pipette was used to extract 128.4 $\mu$l of ethanol and a 10 $\mu$l capacity pipette was used to extract 10 $\mu$l of undiluted cymene to a small, previously marked test tube in order to obtain a 500 mM cymene concentration. A test tube shaker was used to stir the compound until it was homogeneous. A pipette tip was changed prior to being used in a newly obtained dilution to prevent mixing of two different concentrations. In a new test tube, 10 $\mu$l of 500 mM cymene concentration was diluted with 100 $\mu$l of ethanol to acquire 50 mM cymene concentration. Similarly, 10 $\mu$l of each newly obtained cymene concentration was diluted by 100 $\mu$l of ethanol to acquire ten times smaller concentration until 0.005 mM cymene concentration is obtained. Each time the mixture was stirred by avartex shaker and pipette tips were changed prior to handling another cymene concentration.

\begin{figure}
\centering
\includegraphics[width=3in]{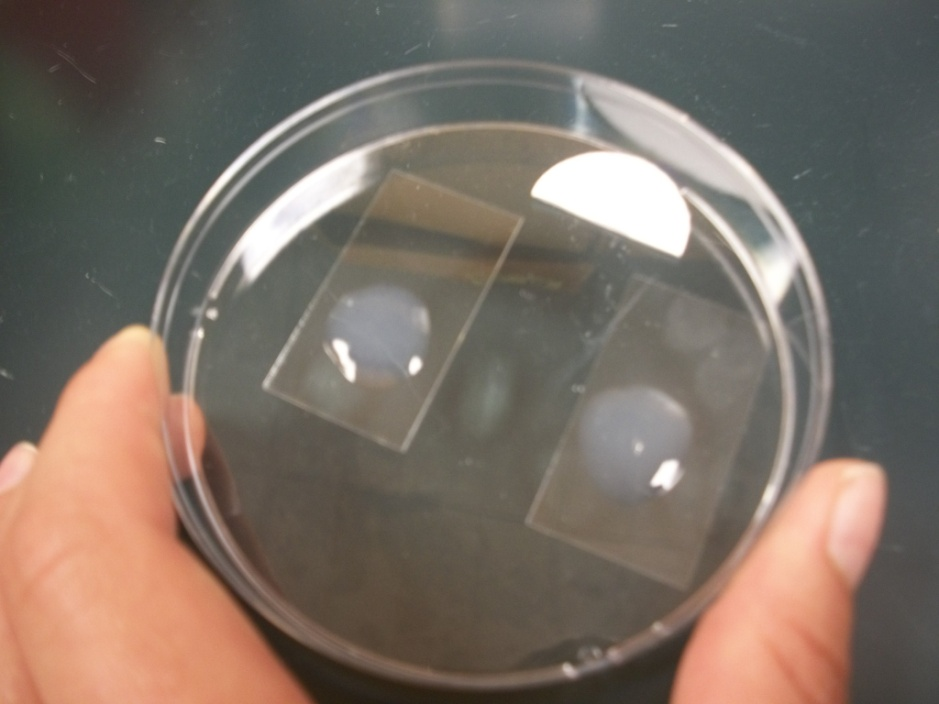}
\caption{Environment for growing \emph{Beauveria bassiana}}
\label{petriDish}
\end{figure}

The environment for the fungal growth was set up in a Petri dish by placing two microscope slides on the opposite sides of the Petri dish and half of the filter glass in front of the microscope slides (Fig. \ref{petriDish}). After 200 $\mu$l of water agar was applied by a pipette to both microscope slides to provide water that is essential for fungal spores to grow, 10 $\mu$l of an oil concentration was applied to the filter glass. A sterilized wire spatula was used to put spores into a test tube filled with water. Since spores tend to cling together in water, a small amount of detergent (Tween – 20) was applied to the dilution to disunite spores. The dilution was sonicated to homogenize mixture. Finally, 10 $\mu$l of the dilution was applied on the top of a water agar on each microscope slide. The Petri dish was closed and placed on a wet sponge in order to keep the spores moist over the time period of the experiment. The procedure was repeated until all the cymene concentrations and the pure ethanol were applied to three Petri-dishes each. The date and cymene concentration were indicated on each Petri dish. The experiment was performed twice.

At certain time points, measurements were done for each cymene concentration and control. The microscope slides were placed under microscope with a camera attached at the top. Multiple slides of spores were captured under the same magnification for each oil concentration. Since the microscope had a calibrated reticle for measuring, a distinct spore, with or without a germination tube, would be chosen and its length would be read from the microscopic ruler. After the images were printed, the chosen spore was found and its length was indicated next to it. Images were separated by concentration and time point; each concentration had a different reference spore of known length. The first data set was obtained at 16, 23, and 40 hours. The second set of measurements was performed at 12, 16, and 23 hours. When images from all concentrations at each hour were obtained and printed, germination tubes were measured by a ruler and entered to Microsoft Excel spreadsheet in millimeters. Data sets, time points, and concentrations were differentiated in the spreadsheet, and the length was converted from millimeters to micrometers. The formula used was:
\begin{equation}\label{convertLength}
1 mm = \frac{L*M}{l}\mu m
\end{equation}
where $L$  represents length of the spore, read from the microscope, $M$ represents the microscope magnification, and $l$ represents length of the same spore in millimeters, measured from the print out. After observing data obtained at 40 hours, it was concluded that the data might not be valid due to cymene oil evaporation. Thus it was agreed upon to discard the data obtained at 40 hours, except for using it to obtain the growth rate value for the model.   
                 
Germination tube length was measured by including both the spore diameter and the germination tube. By observing non-germinated spore lengths over the course of the measurements, it was adjudicated that a non-germinated spore has diameter of 5 $\mu$ms or under. The data was further treated based on this assumption to obtain spore germination percentage for each concentration at each time point.  Absolute percentage germination and percentage germination relative to ethanol were plotted against the log of concentration at each time point in order to observe how percentage germination has changed throughout the time and how each cymene concentration has influenced spore germination compared to ethanol and to other cymene concentrations. Similarly, mean absolute germ tube length and relative germ tube lengths were plotted against log of concentration to determine if some of the cymene concentrations have inhibitory or stimulatory effects on growth of \emph{Beauveria}. This behavior would not be surprising because spores are coated with a strongly hydrophobic layer composed of hydrophobin that repels water molecules. Hence, cymene concentrations might have different impact on the hydrophobic behavior and the water absorption itself. Statistical analyses were performed  using Minitab, Matlab, and R mathematical softwares in order to understand fungal growth behavior in presence of different cymene concentrations and acquire an idea about the possibilities of future modeling strategy.

\subsection{Model assumptions}
Two important assumptions were used in our model of cymene effects on spore germination and fungal growth. The first assumption is that there is an the inversely proportional relationship between the growth rate of germ tube and the lag time period. Snow (1949) assumed that this observation holds true for both a single spore and the whole population \cite{Snow49}. The relationship is demonstrated by the equation:
\begin{equation}
rate * lag = k
\end{equation}
where $k$ is a constant. Dantigny (2007) verified this relationship for populations of eight different species of fungus \cite{Dantigny07}. Our study will test this relationship with individual spores. 

The second assumption of our model is the dilution of water by cymene. Our hypothesis is that to germinate, a spore absorbs water within its vicinity to reach a critical volume and then germination tube starts to form. The time at which it reaches this critical volume is its lag time. The volume of spore at time t is determined by the following equation:
\begin{equation}\label{mainModel}
\frac{d(\frac{V}{V_c})}{dt}=(1+\frac{\gamma x}{\beta + x})(1-\frac{x}{\delta+x})\alpha
\end{equation}

In the formula, $V_c$ is the critical volume and $\alpha$ is the water absorption rate for the control. The quantity x is a theoretical ratio of volume cymene to volume water available in a spore’s vicinity. The relationship between cymene concentrations and ratio x is unknown. However, as cymene concentration decreases ten folds, ratio x also decreases ten folds. This ratio is 0 for control and greater than 0 for the other cymene concentrations. For concentration 500 mM, ratio $x=0.7$ is chosen arbitrary to yield one of the best fit. Further chi-square test can be used to evaluate the goodness of fit. 

\begin{figure}
\centering
\includegraphics[width=5in]{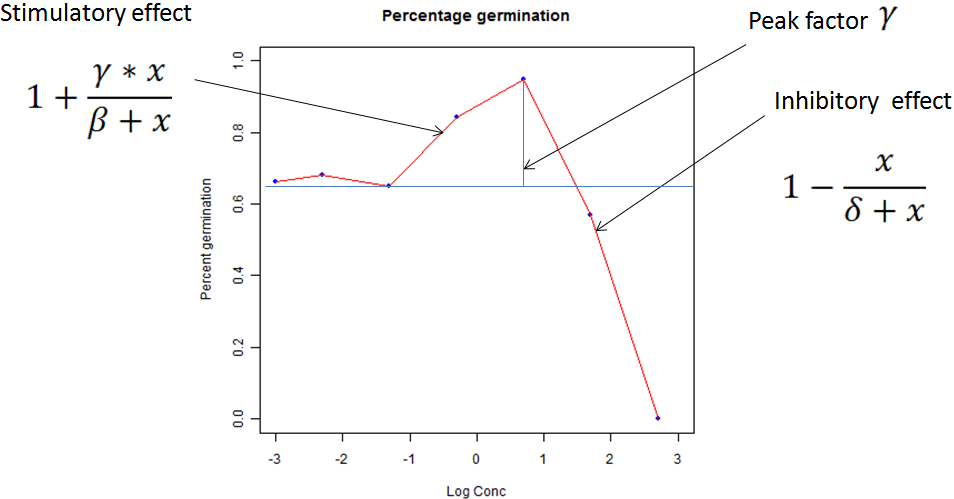}
\caption{Percentage germination was plotted on log cymene concentrations at 16 hours for control and six concentrations of cymene: 0.005 mM, 0.05 mM, 0.5 mM, 5 mM, 50 mM and 500 mM. For control, -3 serves as log of concentration.  From 0 mM (control) to 5 mM, cymene has stimulatory effect on percentage germination. Percentage germination starts to drop at 50 mM,  and is 0 at 500 mM.}
\label{perGerm16}
\end{figure}

A spore starts to germinate when its volume equals the critical volume, i.e. when $\frac{V}{V_c}=1$. At control, $x=0$, thus, water absorption rate is $\alpha = \tau^{-1}(h^{-1})$ where $\tau$ is the lag time of an individual spore. When cymene was added, it had both stimulatory and inhibitory effects on water absorption rates, which in turn influenced lag time and percentage germination. These effects are apparent in the percentage germination data at 16 hours (Fig \ref{perGerm16}). As cymene concentration increases from 0 mM (control) to 5 mM, percentage germination increases. At 50 mM, percentage germination started to drop and there was no germination at 500 mM.

Water absorption rate for an individual spore at different cymene concentration is calculated by:
\begin{equation}
\label{waterAbsEq}
\alpha \prime=(1+\frac{\gamma x}{\beta + x})(1-\frac{x}{\delta+x})\alpha
\end{equation}
where $\alpha \prime$ is water absorption rate corresponding to ratio $x$ and $\alpha$ is the rate at $x=0$. When spore reaches its critical volume at lag time $\tau \prime$, we have $\alpha\prime \tau\prime=1$, and thus:
\begin{equation}
\label{tauCalfromAlp}
\tau \prime=\frac{\tau}{(1+\frac{\gamma x}{\beta + x})(1-\frac{x}{\delta+x})}
\end{equation}
where $\tau$ is lag time at control. As water absorption rate increases, lag time decreases, the spore takes less time to germinate. Therefore, if cymene enhances water absorption rate, it will increase percentage germination as well and vice versa. Water absorption rate and percentage germination vary in the same direction upon changes of ratio $x$. 

In Equation \ref{mainModel}, parameters $\beta$, $\gamma$, $\delta$ are extracted from 16 hour germination percentage data (Fig. \ref{perGerm16}). At concentration 5 mM, the value of percentage germination was highest and was approximately one and a half times the control. This observation was used to estimate the peak factor $\gamma=0.7$. Percentage germination at 0.5 mM is about midway between control and 5 mM. This concentration is corresponding to $x= 0.0007$ and thus $\beta$ is estimated to be 0.001. This value is the concentration at which 50 $\%$ of the population is affected by the stimulatory effect of cymene ($EC_{50}$ value). At 50 mM, percentage germination is approximately midway between the values at 5 mM and 500 mM. Concentration 50 mM is the same as $x$ value 0.07. The parameter $\delta$ is estimated to be 0.1, the $EC_{50}$ of inhibitory effect on spore germination.  

\section{Results}
\subsection{Growth rate under control condition}
In control, no cymene was available. Data were obtained for 12, 16, 23 and 40 hours as described above. From observations, spore diameter averaged about 5 $\mu$ms, ranging from three to seven micrometers. Thus, 5 $\mu$ms was chosen as a threshold for spore diameter. Any measurement under 5 $\mu$ms was considered an ungerminated spore. These spores were not taken into account when growth rate was calculated. For germinated spores, 5 $\mu$ms was subtracted from observed length to obtain net length tube without spore diameter. 
\begin{figure}
\centering
\includegraphics[width=3in]{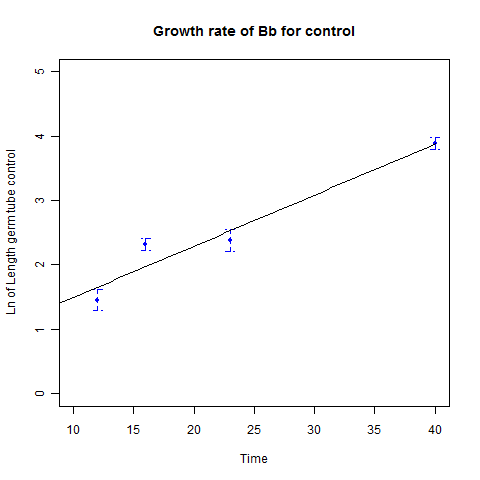}
\caption{Natural log of net tube lengths are plotted against time for control. Each data point is the mean of net tube lengths with corresponding error bars. The fitted curve has slope and intercept estimated by least square linear regression.}
\label{growthCurve}
\end{figure}

Net tube lengths were exponentially proportional to time. Natural logarithm transformation was performed on tube lengths. Log tube lengths increased linearly as time increased. Least square linear regression was used to estimate parameters of the model. The fitted equation is:
\begin{eqnarray}
\label{fittedLengths}
\log\mbox{(net length tube)}&=&0.70186+0.07946*\mbox{time}\\
\mbox{net length tube}&=&2.02e^{(0.07946*\mbox{time})}
\end{eqnarray}

The estimated growth rate was 0.07946 $h^{-1}$ with p-value 0.0299 $(<0.05)$. R square value was 0.9411, which means $94.11\%$ of the variability of net tube lengths can be explained by time, i.e., the goodness of fit is substantial. Fig. \ref{growthCurve}. The intercept value determined the initial value for lengths of germination tube. The initial length value did not have any real biological meaning. At time 0, the initial length is 2.02 ($=e^{0.702}$) $\mu$ms as calculated. However, the model summary indicated that the intercept value 0.70186 is not statistically significant (p-value is $0.1855 >0.05$). Thus, the initial length values will be adjusted later to reach agreement between experimental and theoretical data for the control.

\subsection{Lag time distribution}
\begin{figure}
\centering
\includegraphics[width=3in]{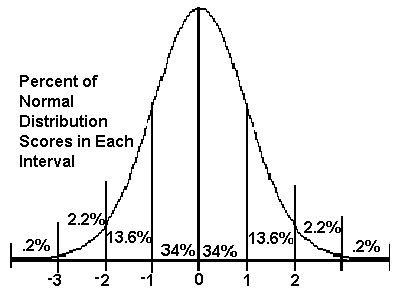}
\caption{The 68-95-99.7 rule of normal distribution (adapted from Google Images)}
\label{normalRule}
\end{figure}

Lag time is defined as the time a spore spent in its inactivated stage. Percentage germination was the ratio of the number of observations with lengths greater than 5 $\mu$ms to the total number of observations. At 4 and 8 hours, no germination was observed. At 40 hours, $96\%$ of observations are greater than 5 $\mu$ms. However, we observed that all spores have germinated at this time. This discrepancy happens because the spore diameters ranged from three to seven micrometers and five was chosen as threshold for the convenience of data treatment. Percentage germination versus time was plotted for the control case at four different time points 12, 16, 23 and 40 hours.
Dantigny et al. (2007) assumed that fungal spore germination lag times have normal distribution. In this research, we also used the same assumption. Mean of lag time distribution was defined as the time at which half of spores have germinated i.e. when the percentage germination was $50\%$. Heuristic estimation of mean and standard deviation of lag time distribution used of the 68-95-99.7 rule of normal distribution. Suppose variable x has normal distribution then about $68\%$ of the values of x are within one standard deviation away from the mean, $95\%$ within two standard deviations and $99.7\%$ within three standard deviations. This rule is illustrated in Fig. \ref{normalRule}.

\begin{figure}
\centering
\includegraphics[width=3in]{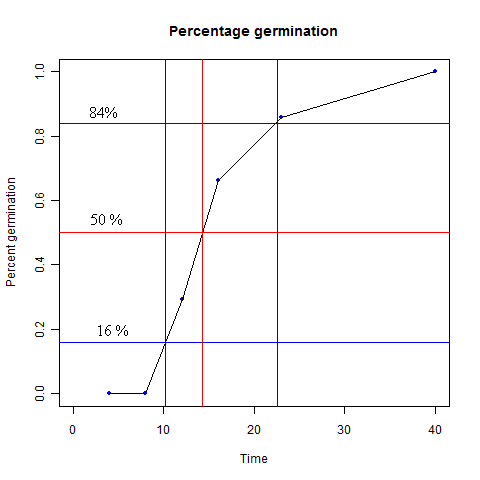}
\caption{The 68-95-99.7 rule was used to estimate mean and standard deviation of lag times. Horizontal lines at 0.16, 0.50 and 0.84 were drawn to estimate the times at which the percentage germination are $16\%, 50\%$ and $84\%$.}
\label{heuristicEstLag}
\end{figure}

From plot of percentage germination versus time, the times at which the percentage germination were $16\%, 50\%$ and $84\%$ were determined (Fig. \ref{heuristicEstLag}). Corresponding times are 10.15, 14.25 and 22.5 hours. Thus, estimated mean was 14.25 hours and estimated standard deviation was 6.125. 

\begin{figure}
\centering
\includegraphics[width=3in]{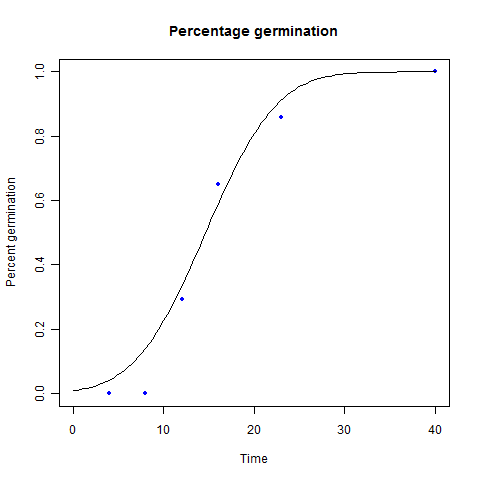}
\caption{Least square nonlinear regression yielded a fitted curve to percentage germination plot. Cumulative distribution function is used as fitted function.}
\label{germPercConFit}
\end{figure}

Non linear regression was performed using R statistical package. The cumulative distribution function of normal distribution was the fitting function. 
\begin{equation}
D(x)=\frac{1}{2}(1+\mbox{erf}(\frac{x-\mu}{\sigma\sqrt{2}}))
\end{equation}
where $\mu$ is the mean, $\sigma$ is the standard deviation and erf is the error function. Least square estimation of the mean was 14.6285 hours and standard deviation was 6.167 hours, very close to the heuristic estimations. The estimated mean was statistically significant. Experimental data and fitted curve for control were shown in Fig. \ref{germPercConFit}. 

\subsection{Lag time and growth rate relationship}
As mentioned in Materials and Methods, a major assumption of this study was the inversely proportional relationship between lag time and growth rate $lag*rate=k$. The product of the mean lag time and the estimated rate found above are used as the constant $k$. This constant does not have a unit. 
\begin{equation}
\label{lagGrowthProd}
lag*rate=k=0.07946h*14.6285h^{-1}=1.162381
\end{equation}
\begin{figure}
\includegraphics[width=5in]{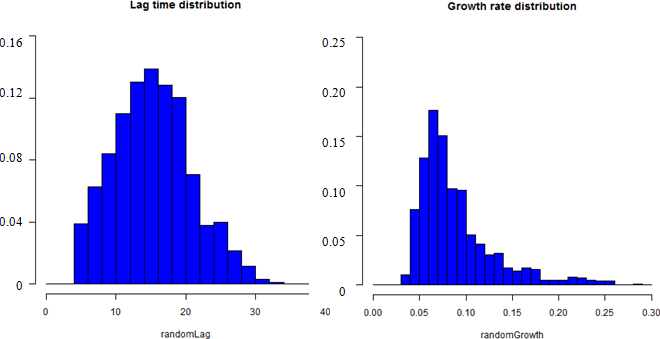}
\caption{(a) Histogram of lag time with normal distribution with mean of 14.6285 hours and standard deviation of 6.167 hours. Cutoff point is 4 hours. (b) Histogram of growth rates calculated from lag time distribution.}
\label{lagNgrowth}
\end{figure}

A thousand lag times were randomly generated from normal distribution with mean of 14.6285 hours and standard deviation of 6.167 hours by R. Some of the lag times generated from this distribution have negative values. This shows that normal distribution inappropriate for lag time distribution. For purpose of this study, we introduced a cutoff point at 4 hours. This cutoff point is chosen arbitrarily to make the simulated data of percentage germination for control agree with experimental data. Each lag time corresponded to a spore. Growth rates were calculated from lag times using the relationship found in Equation \ref{lagGrowthProd}. Distributions of lag times and growth rates for control are presented in Fig. \ref{lagNgrowth}. After growth rates were obtained, lengths of germination tubes were calculated using the formula:
\begin{eqnarray}
\mbox{tube length(t)}&=& L_0 \mbox{  if $t<$ lag}\\
&=& L_0* e^{rate(t-lag)} \mbox{  if $t\geq$ lag}
\end{eqnarray}

\begin{figure}
\centering
\includegraphics[width=3in]{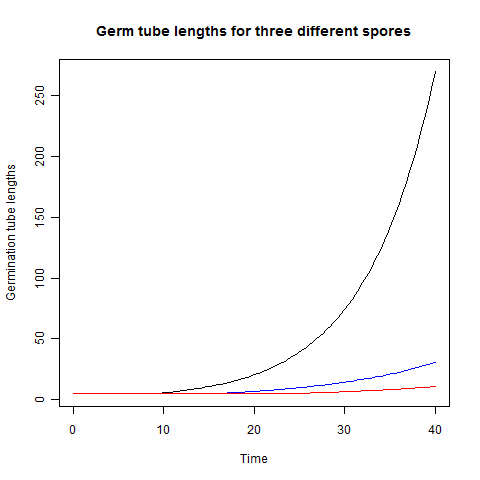}
\caption{Three spores with three different lag times are chosen randomly from the generated populations. Their growth rates were calculated from the relationship with lag times. Three growth curves were presented in the plot.}
\label{3growthCurves}
\end{figure}

\begin{figure}
\centering
\includegraphics[width=5.5in]{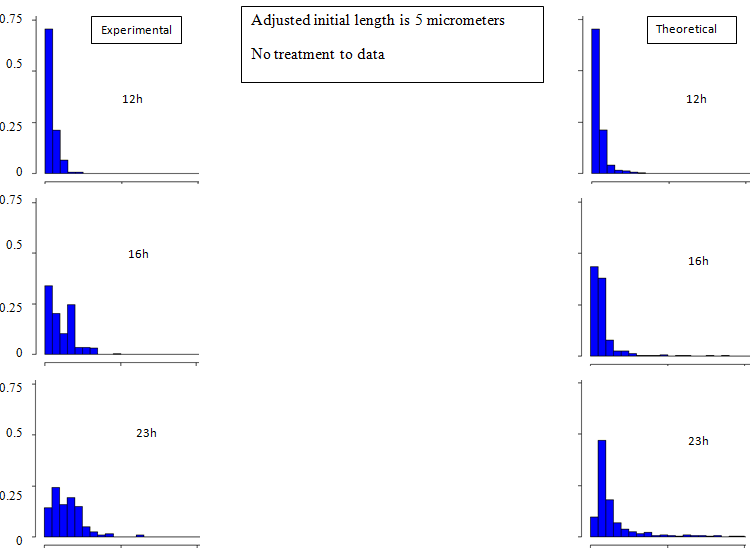}
\caption{ Distribution of germ tube lengths at three experimental time periods for control. No treatment to data was applied. Bin size is 5 $\mu$ms.}
\label{compareCon}
\end{figure}

Initial lengths $L_0$ was chosen to reach agreement between lengths of germination tubes from experiments and from model. In this study, $L_0 = 5\mu$ms  was chosen to yield that agreement. Growth curves of three randomly chosen spores were shown in Fig. \ref{3growthCurves}. Lengths of germination tubes were calculated from these curves. Histograms of tube lengths from experiments and models were shown side by side for comparison in Fig. \ref{compareCon}.Descriptive statistics of these data were presented in Table 1.

\begin{table}\label{descStats}
\caption{Descriptive statistics, comparing theoretical and experimental data for germ tube lengths.}
\centering
\begin{tabular}{|l|l|l|l|l|l|l|} 
\hline
& \multicolumn{2}{|c|}{12 hours} & \multicolumn{2}{|c|}{16 hours}
 & \multicolumn{2}{|c|}{23 hours} \\
\hline
&Expt.&Mode&Expt.&	Model&	Expt.&	Model\\
\hline
Min&0.833&5&1.786&5&2.036&5\\
\hline
Max&23.324&50.67174&45.18&94.16197&63.116&	147.7996\\
\hline
Mean&4.916732&6.414679&11.07827&9.389155&13.99536&15.83705\\
\hline
Std.dev&3.751783&4.168555&7.982262&11.24463&9.794145&19.25989\\
\hline
Median&4.165&5&9.036&5.342344&13.234& 8.849472\\
\hline
Skewness&2.11468&4.641366&1.048792&4.397724&1.584833&3.460279\\
\hline
\end{tabular}
\end{table}

The descriptive statistics and histograms show that the model preserves the positive skewness of experimental data. This means the distribution had long right tail and more density of data on the left. Means and standard deviations also agreed with experimental data. 

Percentage germinations for the model at 12, 16, and 23 hours were calculated. At time t, percentage germination was the ratio of the number of spores with lag times greater than t to the total number of spores. Percentage germinations were compared between experimental and theoretical data in table 2. The fact that theoretical data agreed with experimental data verified the relationship between lag times and growth rates of the fungus \emph{B. bassiana}

\begin{table}\label{comparePerGerm}
\caption{Percentage germination at 12, 16 and 23 hours for experimental and theoretical data.}
\centering
\begin{tabular}{|c|c|c|c|}
\hline
&12h&16h&23h\\
\hline
Experimental&	0.2926829&0.6613546&0.8571429\\
\hline
Theoretical&0.3179433&0.5708290&	0.91290661\\
\hline
\end{tabular}
\end{table}

\subsection{ Simulation of effect of cymene on percentage germination and growth of B. bassiana}
\begin{figure}
\includegraphics[width=5in]{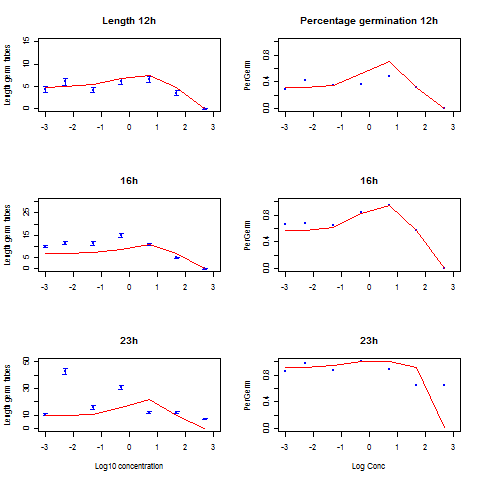}
\caption{Percentage germination and tube length at 12, 16 and 23 hours from experimental and theoretical data plotted side by side to show the fitness of the model.The blue dots are experimental data and the blue lines are theoretical. Tube length plots are shown on the left and percentage germination plots on the right.}
\label{sidebysidePlots2}
\end{figure}

\begin{figure}
\centering
\includegraphics[width=3.5in]{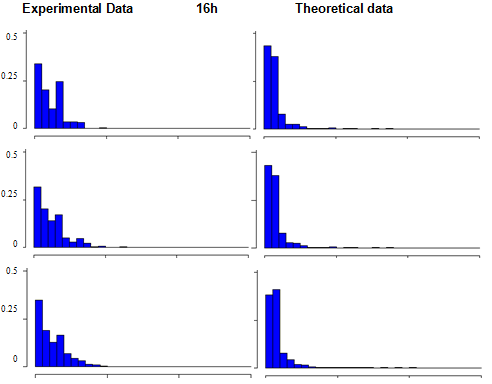}
\includegraphics[width=3.5in]{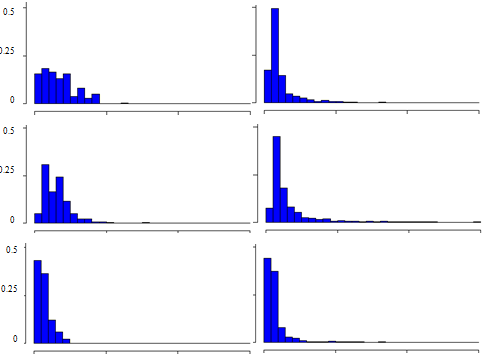}
\includegraphics[width=3.5in]{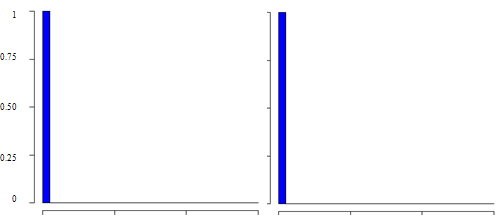}
\caption{Histograms of experimental and theoretical tube lengths (untreated data) plotted side by side for further comparison. Bin size is 5 $\mu$ms. Histograms are arranged in order of increasing cymene concentration. Heights of the bins are probabilities.}
\label{hist16Compare}
\end{figure}

Dilution effect assumption was explained in Material and Method section. Model equation was constructed and parameters were estimated. 
\begin{eqnarray}
\frac{d(\frac{V}{V_c})}{dt}&=&(1+\frac{\gamma x}{\beta + x})(1-\frac{x}{\delta+x})\alpha\\
\beta&=&0.001 \\
\gamma&=&0.7\\
\delta&=&0.1
\end{eqnarray}

The equation and estimations of the three parameters were used to calculating percentage germinations and tube lengths at 12, 16 and 23 hours. Experimental and theoretical values are compared in Fig. \ref{sidebysidePlots2}. In the plot of tube lengths, experimental and theoretical were treated so that only germinated spores were kept. The means of net tube were plotted against log10 of concentrations. Theoretical data fit experimental data. Further statistical analysis is needed for evaluating the goodness of fit. 

Histograms produced by theoretical data preserved some important features of experimental data. Histograms had long right tails. This means large portion of population had short lengths between 0 and 40 $\mu$ms and fewer had lengths over 40 $\mu$ms. As the concentration of cymene increases from 0 to 5 mM, the growth of germination tubes are stimulated by cymene. This is shown in the histograms by the shifting of distribution to the right. At 0.5 and 5 mM, the highest bin was 5-10 $\mu$m instead of 0-5 $\mu$m as in preceding histograms. The widths of histograms decreased at 50 mM. Compared to 5 mM, the width was slightly reduced. At 500 mM, there are only lengths between 0 and 5 $\mu$ms, corresponding to ungerminated spores. 

\section{Discussions}
Smaller concentrations of cymene have a stimulatory effect on growth of \emph{Beauveria bassiana}, while greater concentrations have an inhibitory effect. Similarly, smaller concentrations of cymene attract insects, while greater concentrations of cymene act as insect repellant \cite{Mahaffey04}. \emph{Beauveria bassiana} is an effective entomopathogen, the concentrations that attract the insect may induce germination of the spores, thus allowing the fungus to locate a suitable host.  Cymene is volatile and evaporates in a couple of hours, and thus a seed is protected at initial stage until \emph{B.  bassiana} spores germinate. Germ tubes in our experiment grew in all directions randomly, searching for nutrients that were not provided. If there are no nutrients, the fungus will not survive. Since plants provide food for fungi, fungi have to reach a plant as soon as possible in order to survive. Thus, appropriate cymene concentration should be present that will allow fungi to inhibit the plant and that will also repel, rather than attract insects.

Our study used the assumption that lag times have normal distribution that was demonstrated by Dantigny (2007). This distribution produced negative lag times in the simulation. To avoid this, Dantigny (2007) introduced skewness into normal distribution. Another way is to find a better fitting distribution, ones with all positive values; this will be one of our future directions. The percentage germination over time were used in both Dantigny's paper and this study. While Dantigny used Gomperzt and logistic equation to fit the germination data, we used the cumulative distribution function of normal distribution.

A major difference in the approach used by Dantigny (2007) and the one in this study is that he measured the radial growth of fungal colonies.These data were used to calculate the growth rate of the whole population. This was not possible in our study because \emph{B. bassiana} produced multiple colonies on petri dishes, making it impossible to measure radial growth. We chose therefore to measure germ tube lengths.

A primary assumption for Dantigny's and our study used was that there is an inverse proportionality between growth rate and lag time. While Dantigny tested this assumption on the whole fungal population, our study verified this relationship for the germination of  single spores. 

This research is important because it allows a relatively simple mathematical model of the development of spore germ tubes in the absence and presence of a natural biopesticide.  This model may prove applicable to evaluation of biopesticides.  




\end{document}